\documentclass[12pt]{iopart}
\usepackage{graphicx}
\newcommand{\be}{\begin{equation}}
\newcommand{\ee}{\end{equation}}

\begin{document}

\title[Darboux transformation and generalized harmonic oscillators]{Generalization of
the Darboux transformation and generalized harmonic oscillators}

\author{Dae-Yup Song\dag\ddag $~$ and John R. Klauder\S}
 \address{\dag\ Department of Physics, University of Florida,
Gainesville, FL 32611, USA}
\address{\ddag\ Department of
Physics, Sunchon National University, Suncheon 540-742, Korea }
 \address{\S\ Department of Physics and Mathematics, University of Florida,
Gainesville, FL 32611, USA}
 \eads{\mailto{dsong@sunchon.ac.kr}, \mailto{klauder@phys.ufl.edu}}

\begin{abstract}
The Darbroux transformation is generalized for time-dependent
Hamiltonian systems which include a term linear in momentum and a
time-dependent mass. The formalism for the $N$-fold application of
the transformation is also established, and these formalisms are
applied for a general quadratic system (a generalized harmonic
oscillator) and a quadratic system with an inverse-square
interaction up to $N=2$. Among the new features found, it is
shown, for the general quadratic system, that the shape of
potential difference between the original system and the
transformed system could oscillate according to a classical
solution, which is related to the existence of coherent states in
the system.
\end{abstract}

\section{Introduction}
There has been great interest in using the Darboux transformation
\cite{Darboux,Crum} for the analysis of physical systems and for
finding new solvable systems. It has been shown that the
transformation method is useful in finding soliton solutions of
the integrable systems \cite{MS} and constructing supersymmetric
quantum mechanical systems \cite{SSQM}. An excellent survey for
developments and some applications of the transformation method is
given in \cite{Rosu}.

If a Darboux transformation is applied to a time-independent
Schr\"{o}dinger equation of confining potential, one of the
immediate consequences is that the energy eigenvalues of the
transformed system will be almost identical with those of the
original system except for a finite number of addition(s) and/or
deletion(s) to the spectrum \cite{LP}. On the other hand, Abraham
and Moses (AM) have developed an integral equation algorithm
\cite{AM}, which can also be used to find a new solvable system
based on a known one. When applied to a harmonic oscillator, the
AM algorithm gives a transformed system whose energy spectrum
coincides with that of the harmonic oscillator except that the
lowest eigenvalue has been removed. It has been shown that the
result given by AM can also be derived through the factorization
method \cite{Mielnik} and the factorization method has been
applied for various systems \cite{Fernandez}. After all, the AM
method and the factorization method are intimately related to the
Darboux transformation \cite{LP,SL,BS}.

In order to obtain a new solvable system by implementing the
Darboux transformation, it is necessary to choose the auxiliary
function of the transformation judiciously \cite{LP}. After the
transformation being extended to include a time-dependent
potential, the application of the Darboux transformation and the
two-fold application of the transformation have been explicitly
carried out for the harmonic oscillator model (with time-dependent
frequency), to obtain new solvable systems \cite{BS,Samsonov}. For
the simple harmonic oscillator, unphysical negative-energy
eigenstates, which can be obtained by invoking the symmetry of the
Hamiltonian, have been used as the auxiliary functions of the
Darboux transformation. For the two-fold application of the
transformation, a theorem has already been established in
\cite{Adler} for the choice of the auxiliary functions.

In this article, we will generalize the Darbroux transformation to
be applicable to time-dependent Hamiltonian systems which include
a term linear in momentum and a time-dependent mass. The formalism
for the $N$-fold application of the transformation will also be
established. The transformation method will then be applied for a
general quadratic system (a generalized harmonic oscillator) and a
quadratic system with an inverse-square interaction up to $N=2$.
The general quadratic system \cite{Song99} is known to be related
to a simple harmonic oscillator through unitary transformations,
which may be a reason for the existence of coherent states in the
simple harmonic oscillator \cite{Song}. Even for the cases already
considered \cite{BS,Samsonov}, new features of the Darboux
transformed systems will be given. In particular, it will be shown
that, the shape of the potential difference between the original
Hamiltonian and the transformed one could oscillate according to
the classical solution of the quadratic system, which is related
to the existence of coherent states in the system.

\section{A generalization of the Darboux transformation}

For smooth functions $P(x)$, $Q(x)$, we assume a function $u(x)$
satisfies
 \be
  {d^2 u\over dx^2}+P(x){d u\over dx}+(Q(x)+C)u=0,
 \ee
where $C$ is a constant. If $\phi(x)$ satisfies the linear
differential equation
 \be
 {d^2 \phi\over dx^2}+P(x){d \phi\over dx}+Q(x)\phi=0,
 \ee
Darboux shows that the following equation is true \cite{Darboux}
 \be
\left[{d^2\over dx^2}+P(x){d\over dx} +{dP\over dx}+Q(x)+2({d^2\ln
u\over dx^2})\right]\left({d\phi\over dx}- \phi{d u\over
dx}\right)=0.
 \ee
This Darboux transformation has immediate consequences for a
time-independent Schr\"{o}dinger equation \cite{LP}.

For the extension of the transformation to be applicable for the
time-dependent system described by the Hamiltonian
 \be
 H={p^2 \over 2M(t)}+\left(R(x,t)p+pR(x,t)\right)+V(x,t),
 \ee
we consider the operator
 \begin{eqnarray}\hskip-1cm
 O(t,x)&=&-\rmi\hbar{\partial \over \partial t}+H     \cr
 &=&-\rmi\hbar{\partial \over \partial t}-{\hbar^2\over
 2M(t)} {\partial^2 \over \partial x^2}
 -\rmi\hbar\{2R(x,t){\partial \over \partial x}+ R'(x,t) \}+V(x,t).~~
 \end{eqnarray}
In $H$, $M(t)$ denotes time-dependent mass and the term
proportional to $R(x,t)$ is included to resemble a three
dimensional model which is under a vector potential. A wave
function of the system of $H$ should satisfy the time-dependent
Schr\"{o}dinger equation
 \be
 O(t,x)\psi(t,x)=0.
 \ee
We also introduce the auxiliary functions $u_k$ satisfying
 \be
 O(t,x)u_k (t,x)=0~~~~(k=1,2,\cdots,N).
 \ee
We assume that $u_k$ is a smooth function for any finite $x$, but
$u_k$ need not be square-integrable as it is just an auxiliary
function.

From (6) and (7), it is straightforward to show that
 \be
 \left( O(t,x)-2\rmi\hbar R'(x,t)-{\hbar^2 \over M(t)}
 {\partial^2 \ln u_1 \over \partial x^2} \right)\psi^{1d}(x,t)=0,
 \ee
where
 \be
 \psi^{1d}(x,t)= \left(\psi'-\psi{u_1' \over u_1}\right),
 \ee
and $'$ denotes the partial derivative with respect to $x$.
Equation (8) is a generalization of the Darboux transformation.

The Darboux transformation can be applied repeatedly. For the
description of $k$-fold transformations, we define the Wronskian
determinants as
\begin{eqnarray}
W_k&=\left|
  \begin{array}{cccc}
     u_1&u_2&\cdots&u_k\\
     u_1'&u_2'&\cdots&u_k'\\
     \cdot&\cdot&\cdots&\cdot\\
     \cdot&\cdot&\cdots&\cdot\\
     u_1^{(k-1)}&u_2^{(k-1)}&\cdots&u_k^{(k-1)}
  \end{array} \right|,\cr
 & \cr& \cr
W_{k,\psi}&=\left|
  \begin{array}{ccccc}
     u_1&u_2&\cdots&u_k&\psi\\
     u_1'&u_2'&\cdots&u_k'&\psi'\\
     \cdot&\cdot&\cdots&\cdot&\cdot\\
     \cdot&\cdot&\cdots&\cdot&\cdot\\
     u_1^{(k)}&u_2^{(k)}&\cdots&u_k^{(k)}&\psi^{(k)}
  \end{array} \right|.
\end{eqnarray}
Making use of Crum's formula \cite{Crum}
 \be
 W_{k,\psi}W_{k-1}=W_k{\partial \over \partial x}W_{k-1,\psi}
   -W_{k-1,\psi}{\partial \over \partial x}W_k,
 \ee
one can find that $\psi^{kd}(x,t)$ satisfying
 \be
 \left( O(t,x)-2\rmi k\hbar R'(x,t)-{\hbar^2 \over M(t)}
 {\partial^2 \ln W_k \over \partial x^2} \right)\psi^{kd}(x,t)=0
 \ee
is given by
 \be
 \psi^{kd}(x,t)={W_{k,\psi} \over W_k}.
 \ee

There is a difficulty in interpreting equation (12) as a
Shr\"{o}dinger equation, since the associated Hamiltonian is, in
general, not Hermitian. As discussed in special cases \cite{BS},
this difficulty can be resolved if there exists a purely
time-dependent function $\alpha_k(t)$ satisfying
 \be
 {d \over d t}\ln \alpha_k
 = 2kR'-\rmi{\hbar \over 2M(t)}{\partial^2 \over \partial x^2} \ln {W_k\over
 \bar{W_k}},
 \ee
where $\bar{W_k}$ denotes the complex conjugate of $W_k$. In this
case, equation (12) can be rewritten as
 \be
 \left(O(t,x)-{\hbar^2 \over 2M(t)}
 {\partial^2(W_k\bar{W_k})\over \partial
 x^2}\right)\psi^{kD}(x,t)=0,
 \ee
where
 \be
 \psi^{kD}(x,t)=\alpha_k(t)\psi^{kd}(x,t).
 \ee
If $V(x,t)$ is a real function, equation (15) is the
Schr\"{o}dinger equation of a system described by the Hermitian
Hamiltonian
 \be \hskip-1cm
 H_k={p^2 \over 2M(t)}+\left(R(x,t)p+pR(x,t)\right)+V(x,t)
 -{\hbar^2 \over 2M(t)} {\partial^2 \ln (W_k\bar{W_k}) \over \partial
 x^2}.
 \ee
One of the crucial conditions for $\psi^{kD}(x,t)$ being
square-integrable is that $W_k$ should not have any zero in the
entire space of $x$. Indeed, the derivations of (12) and (13) are
not valid for the zeros of $W_1,~W_2,\cdots,~W_k$, while, if $W_k$
has no zero, the formulas would still be useful even for the cases
that $W_1,~W_2,\cdots,~W_{k-1}$ have zeros, as examples will show
later.

In addition to $\psi^{kD}(x,t)$, other solutions of the
Schr\"{o}dinger equation of $H_k$ would exist \cite{LP}. For the
$H_1$ system, from the fact that
 \be
 \hskip-2cm
 \rmi\hbar{\partial \over \partial t}{1\over u}
 ={\hbar^2 \over 2M}\left({\partial^2 \over \partial x^2}{1\over
 u}\right)
 -2\rmi\hbar R\left({\partial \over \partial x}{1\over u}\right)
 +(-V+\rmi\hbar R'){1\over u}
 +{1\over u}{\hbar^2 \over M} \left({\partial^2 \ln u\over \partial
 x^2}\right),
 \ee
a solution $\psi_u^{1D}$ satisfying $[-i\hbar(\partial/\partial
t)+H_1]\psi_u^{1D}=0$ is given as \cite{LP,BS}
 \be
 \psi_u^{1D}(x,t)={1\over \alpha\bar{u}_1}
 ={1\over\alpha\bar{W}_1}.
 \ee

\section{A general quadratic system}
For an application of the transformation of the previous section,
it may be essential to find $u_1$ or $W_k$ which does not vanish
over the whole coordinate space. Moreover, for an Hermitian
Hamiltonian, such auxiliary solutions are required to support the
existence of a purely time-dependent function, $\alpha_k(t)$,
defined in (14). For the non-vanishing $u_1$ of the harmonic
oscillator, unphysical negative-energy eigenstates can be used as
the auxiliary functions for the transformation. For the two-fold
application, the sign theorem established in \cite{Adler} may be
used to find non-vanishing $W_2$. Nevertheless, it has been shown
that the application and two-fold application of the Darboux
transformation are possible for the harmonic oscillator (of
time-dependent frquency) with and without an inverse-square
potential \cite{BS,Samsonov}.

In this section, we will show that the application and two-fold
application of the Darboux transformation are possible for a
general quadratic system. Even for the cases already considered
\cite{BS,Samsonov}, new features of the applications will be
found.

\subsection{Darboux transformation}
A general quadratic system can be conveniently described by the
Lagrangian \cite{Song99}
 \be
 \hskip-2.5cm
 L_Q = {1\over 2} M(t) \dot{x}^2 - {1\over 2}M(t)w^2(t)
x^2 +F(t) x   + {d \over dt} \left(M(t)a(t) x^2\right) + {d \over
dt} (b(t) x) +f(t),
 \ee
where the overdot denotes the derivative with respect to $t$. This
Lagrangian gives the classical equation of motion
 \be
 {d \over {dt}} \left(M(t) \dot{{x}}\right) + M(t) w^2(t) {x} =F(t)
 \ee
which is that of a generalized harmonic oscillator of mass $M(t)$
and frequency $w(t)$ in an external force $F(t)$. The
corresponding Hamiltonian is written as
 \be
 \hskip-2.5cm
  H_Q= {{p}^2 \over 2 M(t)} - a(t)[{p}{x}+{x}{p}]
     +{1\over 2} M(t)c(t){x}^2
     -{b(t)\over M(t)} {p}+d(t){x}+( {b^2(t) \over 2M(t)} -f(t)),
 \ee
where
 \be
 c(t)=w^2 + 4a^2 -2 \dot{a} -2 {\dot{M}\over M}a,~~~
 d(t)= 2ab-\dot{b} -F.
 \ee
We assume that $H_Q$ is Hermitian. The general solution of
equation (21) is a linear combination of a particular solution
$x_p(t)$ and two linearly independent homogeneous solutions
$u(t),v(t)$. We assume $x_p(t),u(t),v(t)$ are real, and define
$\rho(t)$ and a time-constant $\Omega$, for later use, as
 \be
 \rho=\sqrt{u^2(t)+v^2(t)}, ~~ \Omega= M(t)[ \dot{v}(t)u(t) - \dot{u}(t)v(t)].
 \ee

Since the general quadratic system can be obtained from a simple
harmonic oscillator through unitary transformations, we first
consider the simple harmonic oscillator of unit mass and
frequency. The Schr\"{o}dinger equation for the simple harmonic
oscillator, whose time is $\tau$, is written as
 \be
 \rmi\hbar{\partial \over \partial \tau}\psi_s =
 -{\hbar^2\over 2}{\partial^2 \over \partial x^2}\psi_s
 +{x^2\over 2}\psi_s= H_s\psi_s.
 \ee
Making use of the invariance of the Schr\"{o}dinger equation under
the exchange of $\tau\leftrightarrow -\tau$ and $x\leftrightarrow
ix$, from the well-known wave functions
 \be
\psi_n^s(\tau,x)={\rme^{-\rmi(n+1/2)\tau}\over (2^n
n!\sqrt{\pi\hbar})^{1/2}} \exp\left[-{x^2\over
2\hbar}\right]H_n\left({x\over \sqrt{\hbar}}\right)
 \ee
one can find auxiliary functions satisfying the same
Schr\"{o}dinger equation as
 \be
 v_n^s(\tau,x)=\rme^{\rmi(n+1/2)\tau}
 \exp\left[{x^2\over 2\hbar}\right]H_n\left({\rmi x\over
 \sqrt{\hbar}}\right),~~~~ n=0,1,2,\cdots.
 \ee
Indeed, $v_n^s(\tau,x)$ for even $n$ has no zero over the whole
coordinate space, and it can be used as an auxiliary function of
the Darboux transformation. In this subsection, we restrict our
attention on the cases that $n=0,2,4,\ldots$.

By defining
 \be O_s(\tau,x)=-\rmi\hbar{\partial\over \partial \tau} +H_s,~~~
 O_Q(t,x)=-\rmi\hbar{\partial\over \partial t} +H_Q,
 \ee
and if $\tau$ and $t$ are related by
 \be d\tau={\Omega \over M(t)\rho^2(t)} dt,
 \ee
then from the results given in \cite{Song} one can find that
 \be
 O_Q(t,x)=U_GU_FU_SO_s(\tau,x)U_S^\dagger U_F^\dagger U_G^\dagger|_{\tau=\tau(t)},
 \ee
where
\begin{eqnarray}
U_G &=&  \exp[{\rmi\over \hbar}(M(t)a(t)x^2+b(t)x +\int^tf(z)dz)]  \\
U_F(x_p) &=& \exp[{\rmi\over \hbar}(M\dot{x}_px+\delta(x_p))]
          \exp(-{\rmi\over \hbar}x_p p)    \\
U_S(\rho,\Omega) &=& \exp[{\rmi \over 2 \hbar}M {\dot{\rho} \over
\rho}x^2] \exp[-{\rmi\over 4\hbar}\ln({\rho^2 \over
\Omega})(xp+px)]
\end{eqnarray}
with $\delta$ defined through the relation
 \be \dot{\delta}(x_p)={1\over 2}Mw^2x_p^2-{1\over 2}M\dot{x}_p^2.
 \ee

A point that should be mentioned is the unitary operators $U_F$
and $U_S$ are not unique since they depend on the choice of
classical solutions. Instead of $\{u,v\}$ and $x_p$, one can take
another set of two linearly independent homogeneous solutions
$\{\tilde{u},\tilde{v}\}$ and a particular solution $\tilde{x}_p$
of equation (21). After defining $\tilde{\rho}, \tilde{\Omega}$
and $\tilde{\delta}$ from $\{\tilde{u},\tilde{v}\}$ and
$\tilde{x}_p$ as $\rho, \Omega$ and $\delta$ are defined from
$\{u,v\}$ and $x_p$, one can find that the unitary relation (30)
is also valid with the unitary operators
$U_F(\tilde{x}_p),~U_S(\tilde{\rho},\tilde{\Omega})$.

Making use of the unitary relation, the normalized wave functions
of the system of $H_Q$ have been given in \cite{Song} as
\begin{eqnarray}
\hskip-2cm \psi_m^Q(t,x)&=& {1\over \sqrt{2^m m!}}\left({\Omega
\over \pi\hbar}\right)^{1\over 4}
     {1\over \sqrt{\rho(t)}}\left[{u(t)-\rmi v(t) \over \rho(t)}
     \right]^{m+{1\over 2}}
     \exp\left[{\rmi\over\hbar}\left(\delta(t) +\int^t f(z)dz\right)\right]
\cr&&\times
     \exp\left[{\rmi\over\hbar}\left[M(t)a(t)x^2+ (M(t)\dot{x}_p(t)+b(t))x
       \right]\right]
\cr&&\times
     \exp{\left[{(x-x_p(t))^2\over 2\hbar}\left(-{\Omega \over \rho^2(t)}
               +\rmi M(t){\dot{\rho}(t) \over \rho(t)}\right)\right]}
\cr&&\times
         H_m\left(\sqrt{\Omega \over \hbar} {x -x_p(t) \over
         \rho(t)}\right).~
\end{eqnarray}
The unitary relation can also be used to find a solution
$v_n^Q(t,x)$ of the Schr\"{o}dinger equation
 \be
 O(t,x)v_n^Q(t,x)=0,
 \ee
as
 \begin{eqnarray}
\hskip-2cm v_n^Q(t,x)
 &=&U_GU_F(\tilde{x}_p)U_S(\tilde{\rho},\tilde{\Omega})
     v_n^s(\tau,x)|_{\tau=\tau(t)}\cr
 &=&\sqrt{\sqrt{\tilde\Omega}\over \tilde\rho(t)}
   \left[{\tilde u(t)+\rmi\tilde v(t) \over \tilde\rho(t)}\right]^{n+{1\over 2}}
     \exp\left[{\rmi\over\hbar}\left(\tilde\delta(t) +\int^t f(z)dz\right)\right]
\cr&&\times
     \exp\left[{\rmi\over\hbar}\left[M(t)a(t)x^2+
     \left(M(t)\dot{\tilde x}_p(t)+b(t)\right)x\right]\right]
\cr&&\times
     \exp{\left[{(x-\tilde x_p(t))^2\over 2\hbar}
          \left({\tilde\Omega \over \tilde\rho^2(t)}
               +\rmi M(t){\dot{\tilde\rho}(t) \over \tilde\rho(t)}\right)\right]}
         H_n\left(\rmi\sqrt{\tilde\Omega \over \hbar}
         {x -\tilde x_p(t) \over \tilde\rho(t)}\right).
\end{eqnarray}
From the properties of the unitary transformations, it is manifest
that $v_n^Q(t,x)$ for even $n$ does not have a zero over whole
coordinate space.

Since, for even $n$, $H_n\left(\rmi\sqrt{\tilde \Omega \over
\hbar} {x -\tilde x_p(t) \over \tilde\rho(t)}\right)$ is a real
function, it is easy to find that
 \be
 -\rmi{\hbar \over 2M}{\partial^2 \over \partial x^2}\ln {v_n^Q\over \bar{v}_n^Q}=
2a+{\dot{\tilde\rho}\over\tilde\rho},
 \ee
which shows, for the choice of $u_1^Q=W_1^Q=v_n^Q$, that
$\alpha_1^{Q}(t)$ defined in (14) can be found as
 \be
 \alpha_1^Q(t)=\tilde\rho(t)
 \ee
up to a normalization constant. Therefore, the transformation
formalism developed in the previous section can be applied with
$u_1^Q(t,x)= v_n^Q(t,x)$ to give the solvable model described by
the Hamiltonian
 \be \hskip-2cm
 H_1^{nQ}(t,x,p)=H_Q-{\hbar\tilde\Omega \over M\tilde\rho^2}
 +4n{\hbar\tilde\Omega \over M\tilde\rho^2}
 \left[(n-1){H_{n-2}(z) \over H_n(z)}
    -n\left(H_{n-1}(z) \over H_n(z)\right)^2\right].
 \ee
where $z=\rmi\sqrt{\tilde\Omega\over \hbar}{x-\tilde x_p \over
\tilde\rho}$. If we adopt the notation that
$H_{-2}(z)=H_{-1}(z)=0$, equation (40) is valid for
$n=0,2,4,\cdots$. The magnitude of $\Delta V$ $(\equiv H_1^{nQ}
-H_Q+{\hbar\tilde\Omega \over M\tilde\rho^2})$ is $O(\hbar)$.
Since $\Delta V$ vanishes in the limit of $|z|\rightarrow \infty$,
$\Delta V/\hbar$ approaches 0 except for the region in the
vicinity of $x_p$ where the width of the region is
$O(\sqrt{\hbar})$.

From equations (10), (13) and (16), the unnormalized wave
functions $\psi_m^{nQ}(t,x)$ satisfying the Schr\"{o}dinger
equation
 \be
 \left( -\rmi\hbar{\partial \over \partial t}+H_1^{nQ} \right)
 \psi_m^{nQ}(t,x)=0,
 \ee
can be given as
 \begin{eqnarray}
 \hskip-2.5cm
 \psi_m^{nQ}(t,x)=&\tilde{\rho}\psi_m
   [{\rmi\over \hbar}M(\dot x_p-\dot{\tilde x}_p)
     +{x-x_p \over \hbar}\left(-{\Omega \over \rho^2}
     +\rmi M{\dot{\rho}\over \rho}\right)
     -{x-\tilde x_p \over \hbar}
     \left({\tilde\Omega \over \tilde\rho^2}
     +\rmi M{\dot{\tilde\rho}\over \tilde\rho}\right)\cr
 & ~~~~~~~~~
     +2m\sqrt{\Omega\over \hbar}{1\over \rho}{H_{m-1}(w)\over H_m(w)}
     -2\rmi n\sqrt{\tilde\Omega\over \hbar}{1\over \tilde\rho}
     {H_{n-1}(z)\over H_n(z)}],
 \end{eqnarray}
where $w=\sqrt{\Omega\over \hbar}{x-x_p \over \rho}$. As in the
general quadratic system \cite{Song99,Song}, a different choice of
the classical solutions $u(t),v(t)$ $x_p(t)$ gives different wave
functions. When we choose $u(t)=\tilde{u}(t),~v(t)=\tilde{v}(t),$
and $x_p(t)=\tilde x_p(t)$, for even integer $n$, an unnormalized
$\psi_m^{nQ}(t,x)$ is written as
 \be
 \psi_m^{nQ}(t,x)=-\sqrt{2\Omega \over \hbar}\left[\sqrt{m+1}
 \left({u+\rmi v \over \rho}\right)\psi_{m+1}
 +\sqrt{2}n\rmi{H_{n-1}(\rmi w) \over H_n(\rmi w)}\psi_m\right],
 \ee
for $m=0,1,2,\ldots$. For the case of $M(t)=1$ and $a(t)=0$,
$\psi_m^{2Q}(t,x)$ in equation (43) reproduces the wave functions
given in \cite{Samsonov}, up to normalization. From equation (19),
another wave function satisfying $\left(-\rmi\hbar{\partial \over
\partial t}+H_1^{nQ}\right) \psi_{-n}^{nQ}=0$ is given as
 \be
 \psi_{-n}^{nQ}={1\over \rho \bar{u}_n^{Q}}.
 \ee
For non-negative even integer $n$, it is clear that
$\psi_{m}^{nQ}$ $(m=-n,0,1,2\ldots)$ are square-integrable.

Even for the simple harmonic oscillator of unit mass and
frequency, since the shape of the probability density of a wave
function could breathe and oscillate \cite{Song99,Song}, the
region, where $\Delta V/\hbar$ is different from 0 by a certain
amount, could breathe and oscillate. If we take the classical
solutions as $\tilde{u}(t)=u(t)=\cos t,$ $\tilde{v}(t)=v(t)=c\sin
t$ $(c\neq0)$, and $x_p(t)=d\cos t$, the breathing and/or
oscillating behavior of the non-vanishing region appears when
$c\neq 1$ and/or $d\neq 0$, respectively. For the choice $c=1$ and
$d=0$, $H_1^{nQ}$ becomes
 \be
 H_1^{ns}={1\over 2}(p^2+x^2)-\hbar
 -\hbar^2{\partial^2 \over \partial x^2}\ln
 H_n(\rmi{x\over\sqrt{\hbar}}).
 \ee
Through the same choice of classical solutions, for non-negative
even integer $n$, one can also find the eigenfunctions of
$H_1^{ns}$ as
 \begin{eqnarray}
 \psi_m^{s,n}&=&-\sqrt{2\over \hbar}
   \left[\sqrt{m+1}\rme^{\rmi t}\psi_{m+1}^s
   +\sqrt{2}ni{H_{n-1}(\rmi{x\over \sqrt{\hbar}})\over H_n(\rmi{x\over
   \sqrt{\hbar}})}\psi_m^s \right]\cr
 \psi_{-n}^{s,n}&=&{1\over \bar u_n^s},
 \end{eqnarray}
whose eigenvalues are $(m+{1\over2})\hbar$ and
$-(n+{1\over2})\hbar$, respectively, as expected in \cite{LP}.

\subsection{A two-fold transformation}
In this subsection, we will show that the two-fold application of
the Darboux transformation is also possible for a general
quadratic system. It will also be manifested that, in the two-fold
transformation, the breathing and oscillating behavior would still
appear in the transformed systems of a simple harmonic oscillator.

By taking
 \be
 u_1(t,x)=\psi_n(t,x), ~~~u_2(t,x)=\psi_{n+1}(t,x),
 \ee
one can find that
 \begin{eqnarray}
 W_2 &=& \psi_n\psi_{n+1}'-\psi_{n+1}\psi_n' \cr
  &&={1\over 2^n n!\sqrt{2(n+1)}}
     \left({u-\rmi v \over \rho}\right)^{2n+1}
     {1\over \rho}\sqrt{\Omega \over \hbar}\psi_0^2(t,x) J_n(w),
 \end{eqnarray}
where
 \be
 J_n(w)=H_n(w){d\over dw}H_{n+1}(w)-H_{n+1}(w){d\over dw}H_{n}(w)
 \ee
with $w=\sqrt{\Omega\over \hbar}{x-x_p \over \rho}$. Making use of
a recursive relation among the Hermite polynomials \cite{GR}, one
can show that $J_n(w)=2H_n^2(w)+2nJ_{n-1}(w)$. This relation, with
the fact $J_0=2$, shows that $J_n(w)$ is positive definite for all
$x$, which in fact has been implied in \cite{Adler}. One can also
find that
 \be
 -\rmi{\hbar \over 2M}{\partial^2 \over \partial x^2}
 \ln {W_2 \over \bar{W}_2}=4a+ 2{\dot{\rho} \over \rho}.
 \ee
Equation (50) shows that $\alpha_2^Q$ can be given in accordance
with equation (14) as
 \be
\alpha_2^Q(t)=\rho^2(t),
 \ee
up to a multiplicative constant.

With the $W_2$, the two-fold transformation can, therefore, be
applied to the quadratic system to give the transformed
Hamiltonian
 \be
H_2^{nQ}= H_Q+2{\hbar \Omega \over M\rho^2}
      -{\hbar^2\over M}{\partial \over \partial x^2}\ln J_n(w).
 \ee
For the case of $M(t)=1$ and $a(t)=0$, $H_2^{nQ}$ reduces to the
Hamiltonian found in \cite{Samsonov}. Making use of equations
(10), (13) and (16), it is also possible to find the wave
functions of the system of $H_2^{nQ}$. For non-vanishing $W_2$, it
turns out that the same set of classical solutions must be used
for both $u_1$ and $u_2$ in equation (47). However, in obtaining
wave functions, a different set of classical solution could be
used for $\psi$ in equation (10), as in the Darboux transformation
of the model.

\section{A quadratic system with an inverse-square interaction}
In this section, we will consider the application of the Darboux
transformation for the quadratic system with an inverse-square
interaction described by the Hamiltonian
 \be H_{in}={p^2\over 2M(t)}-a(t)(xp+px) +{1\over 2}M(t)c(t)x^2
 +{g\over M(t)}{1\over x^2},
 \ee
defined on the half line $x>0$, where $g$ is a constant. The
system of $H_{in}$ is related to the system described by the
Hamitonian
 \[ H_{in}^s={p^2\over 2} +{1\over 2}x^2 +{g\over x^2},
 \]
through a unitary relation \cite{Song}. If $u(t),v(t)$ denote the
homogeneous solutions of equation (21) as in the previous section,
and $\rho(t)$ and $\Omega$ are defined by equation (24), a wave
function of the system of $H_{in}$ is given, for non-negative
integer $n$, as
\begin{eqnarray}\hskip-2cm
\psi_n^{in}(t,x)&=&\left({4\Omega\over\hbar\rho^2}\right)^{1/4}
\left({\Gamma(n+1) \over \Gamma(n+\alpha+1}\right)^{1/2}
\left({u-\rmi v\over \rho}\right)^{(2n+\alpha+1)} \left({\Omega
x^2\over \hbar\rho^2}\right)^{(2\alpha+1)/4}    \cr &&\times
\exp\left[-{x^2 \over 2\hbar}\left({\Omega \over \rho^2}-\rmi
M{\dot{\rho}\over \rho}-2\rmi Ma\right)\right]
 L_n^\alpha\left({\Omega x^2 \over \hbar\rho^2}\right),
\end{eqnarray}
where $\alpha$ is defined through the relation $g={1\over
2}(\alpha+{1\over 2})(\alpha-{1\over 2})\hbar^2$, and $L_n^\alpha$
is the Laguerre polynomial \cite{GR}. For $\alpha>-1$,
$\psi_n^{in}(t,x)$ is square-integrable on the half line.

Through similar procedures used in the previous section, one can
find the auxiliary function
 \begin{eqnarray}
 v_n^{in}&=&{1\over \sqrt{\rho}}\left({u+\rmi v\over \rho}\right)^{(2n+\alpha+1)}
     \left({\Omega_0 x^2 \over \hbar\rho^2}\right)^{(2\alpha+1)/4}    \cr
&&\times \exp\left[{x^2 \over 2\hbar}\left({\Omega_0 \over
\rho^2}+\rmi M{\dot{\rho}\over \rho}+2\rmi Ma\right)\right]
 L_n^\alpha\left(-{\Omega_0 x^2 \over \hbar\rho^2}\right),
 \end{eqnarray}
satisfying
 \be
 \left(-\rmi\hbar{\partial \over \partial t}+H_{in}\right)v_n^{in}=0.
 \ee
From the properties of the Laguerre polynomial, $v_n^{in}$ has no
zero on the half line, and it is easy to see that
 \be
 -\rmi{\hbar\over 2M}{\partial^2 \over \partial x^2}
 \ln {v_n^{in}\over \bar{v}_n^{in}}=2a+{\dot{\rho}\over \rho}.
 \ee
Equations (14) and (57) show that $\alpha_1^{in}(t)$ can be given
as
 \be
 \alpha_1^{in}(t)=\rho(t)
 \ee
up to a normalization constant. The Darboux transformation can
therefore be carried out, with $u_1=v_n^{in}(t,x)$, to find the
Hamiltonian of a solvable model as
 \be
 H_1^{n,in}=H_{in}-{\hbar^2 \over
 2M}{\partial^2(v_n^{in}\bar v_n^{in}) \over \partial x^2}.
 \ee

In implementing the formulas (10), (13) and (16), different sets
of homogeneous solutions of equation (21) can be used for
$v_n^{in}(=u_1)$ and $\psi_m^{in}(=\psi)$ to find the general
expression of the wave functions of the $H_1^{n,in}$ system, as in
the general quadratic system of the previous section. For
simplicity, however, we only consider the case that the same set
$\{u(t),v(t)\}$ of the homogeneous solutions is used in $v_n^{in}$
and $\psi_m^{in}$. The wave functions $\psi_{m}^{n,in}$ of the
$H_1^{n,in}$ system, if we adopt the notation $L_{-1}(y)=0$, is
given as
 \begin{eqnarray}
 \psi_{m}^{n,in} &=& -2{\rho \over x}\sqrt{m(m+\alpha)}
 \left({u-\rmi v \over \rho}\right)^2 \psi_{m-1}^{in}\cr
 &&+2{\rho \over x}\left[(m-n-1)y+(n+\alpha)
 {L_{n-1}^\alpha(-y) \over L_n^\alpha(-y)}\right]\psi_m^{in},
 \end{eqnarray}
for $m=1,2,3,\ldots$, where
 \be y={\Omega x^2 \over \hbar \rho^2}.\ee
For $m=0$, the wave function of the $H_1^{n,in}$ system is given
as
 \be
 \psi_{0}^{n,in}=2{\rho \over x}\left[-(n+1)y+(n+\alpha)
 {L_{n-1}^\alpha(-y) \over L_n^\alpha(-y)}\right]\psi_0.
 \ee
It is clear that $\psi_{m}^{n,in}(t,x)$ is square-integrable on
the half line for $\alpha>0$. Though another formal solution of
the Schr\"{o}dinger equation of the $H_1^{n,in}$ system can be
found through equation (19), that solution turns out to be not
square-integrable for $\alpha>0$.

As implied in \cite{Adler}, for the two-fold application of the
Darboux transformation, the auxiliary functions can be chosen as
 \be
 u_1(t,x)=\psi_n^{in}(t,x), ~~~u_2(t,x)=\psi_{n+1}^{in}(t,x),
 \ee
which gives
 \be
 W_2^{in}=2{n!\over (n+\alpha)!}\sqrt{n+1 \over n+1+\alpha}
  \left({u-\rmi v \over \rho}\right)^{4n+2}\left(\psi_0^{in}\right)^2
  {\Omega x\over \hbar \rho^2} K_n(y),
 \ee
where
 \be
 K_n(y)=L_n^\alpha(y){d\over dy} L_{n+1}^\alpha(y)-
    L_{n+1}^\alpha(y){d\over dy} L_{n}^\alpha(y).
 \ee
Making use of a recurrence relation among the Laguerre
polynomials, one can easily find that
 \be
 K_n(y)=-{1\over n+1}\left(L_n^\alpha(y)\right)^2
       +{n+\alpha \over n+1}K_{n-1} (y).
 \ee
With the fact that $K_0=-1$, (66) proves that $K_n(y)$ is
negative-definite for all $n$, so that $W_2^{in}$ does not vanish
for all $0<x<\infty$. The fact
 \be
 -\rmi{\hbar \over 2M}{\partial^2 \over \partial x^2}
 \ln {W_2^{in} \over \bar{W}_2^{in}}=4a+ 2{\dot{\rho} \over \rho},
 \ee
shows that the Hermiticity condition (14) of the Hamiltonian can
be satisfied with
 \be
 \alpha_2^{in}(t)=\rho^2(t).
 \ee
With $W_2^{in}$, from (17), the Hamiltonian of the transformed
system can be found as
 \be
 H_2^{in}=H_{in}+{2\hbar^2(\alpha+1) \over M x^2}
    +2{\hbar\Omega\over M\rho^2}
    -{\hbar^2 \over M}{\partial^2 \over \partial x^2}\ln K_n(y).
 \ee
It may be possible, through the formulas (10), (13) and (16), to
find the wave functions of the system of $H_2^{in}$.

\section{Discussion}

We have generalized the Darboux transformation to be applicable to
a general one-dimensional time-dependent Hamiltonian system. The
formalism for an $N$-fold application of the transformation has
also been established. It has been shown that an Hermitian system
can be found from the transformed system, if the auxiliary
function(s) of the transformation could support the existence of a
purely time-dependent function satisfying a certain condition
(equation (14)). The formalisms have been applied to a general
quadratic system and a quadratic system with an inverse-square
interaction. As the potential difference between the original and
the transformed systems is calculated from a (formal) solution of
the Schr\"{o}dinger equation, due to unitary relations responsible
for the existence of coherent states in the system, the shape of
the potential difference could oscillate according to a classical
solution for the general quadratic system.

Since the potential difference is calculated from a solution of
the Schr\"{o}dinger equation, the difference depends on the Planck
constant $\hbar$. In the cases considered, the magnitude of the
potential difference is $O(\hbar)$. The range within which the
potential difference effectively depends on the space coordinate
would also be described in terms of $\hbar$, as has been shown
explicitly in the examples considered. These features are not
usual in the standard text book of quantum mechanics, and the
potential difference appears as if it is a quantum correction
which depends on the position.

It should also be mentioned that, for the systems considered in
this paper, there  may exist other choices of auxiliary functions
which lead to Hermitian systems. Only for limited cases, however,
it may be possible to find exact solutions of the Schr\"{o}dinger
equation of a time-dependent system. Even if the solutions are
found, the auxiliary function should be chosen judiciously so that
the transformed system could be Hermitian. As an example,
$v_1^Q+v_2^Q$ is a solution of the Schr\"{o}dinger equation of the
general quadratic system, so that the Darboux transformation can
be formally applied to it. However, with this solution, one can
not find $\alpha(t)$ satisfying (14).

\ack One of us (DYS) is grateful to Prof. J.H. Park for her help
on Crum's formula. This work was supported in part by the Korea
Research Foundation Grant (KRF-2002-013-D00025)and by NSF Grant
1614503-12.

\end{document}